\documentclass{article}
\usepackage{amsfonts}
\usepackage{amsmath}
\usepackage{epsf}
\usepackage{citesort}

\begin{document}

\title{Multicomponent multisublattice alloys, nonconfigurational entropy and other
additions to the Alloy Theoretic Automated Toolkit}
\author{A. van de Walle}

\maketitle

\begin{abstract}
A number of new functionalities have been added to the Alloy Theoretic
Automated Toolkit (ATAT) since it was last reviewed in this journal in 2002.
ATAT can now handle multicomponent multisublattice alloy systems,
nonconfigurational sources of entropy (e.g. vibrational and electronic
entropy), Special Quasirandom Structures (SQS) generation, tensorial cluster
expansion construction and includes interfaces for multiple atomistic or ab
initio codes. This paper presents an overview of these features geared
towards the practical use of the code. The extensions to the cluster
expansion formalism needed to cover multicomponent multisublattice alloys are
also formally demonstrated.
\end{abstract}

\section{Introduction}

The Alloy Theoretic Automated Toolkit (ATAT) \cite%
{avdw:atat,avdw:maps,avdw:emc2} is a suite of software tools facilitating
the determination of thermodynamic properties of solid state alloys from
first-principles calculations. It relies upon the cluster expansion
formalism \cite%
{sanchez:cexp,ducastelle:book,fontaine:clusapp,zunger:NATO,zunger:scord,wolverton:srorev,ceder:oxides,asta:fppheq}
to build a simplified effective Hamiltonian that accurately reproduces
quantum mechanical calculation results for an alloy system of interest and
that can be used to efficiently calculate their thermodynamic properties.
ATAT is freely distributed and open-source, thus encouraging users to
contribute \cite{arroyave:open}.

ATAT's basic functionalities have been described in an earlier issue of the
CALPHAD journal \cite{avdw:atat}. Since then, a number of new features have
been added and the purpose of the present paper is to describe these
additions and provide a concise \textquotedblleft user
guide\textquotedblright\ for these new features. The most significant
additions to ATAT are

\begin{enumerate}
\item extensions to multicomponent multisublattice systems;

\item the inclusion of nonconfigurational sources of entropy (vibrational
and electronic entropy);

\item Special Quasirandom Structures (SQS) generation;

\item tensorial cluster expansion construction;

\item support for multiple atomistic (ab initio) codes;

\item various new conversion and analysis utilities.
\end{enumerate}

The features presented here are not exhaustive and new features are
continuously being added. Hence, the reader is invited to consult the manual
supplied with the ATAT distribution or posted on the ATAT web site ( \texttt{%
http://alum.mit.edu/www/avdw/atat} ) for the most up to date information.

\section{Multicomponent multisublattice systems}

\subsection{The cluster expansion in multicomponent multisublattice systems}

The traditional \emph{cluster expansion} represents the relationship $%
q\left( \sigma \right) $ between a \emph{configuration} $\sigma $\ and some
scalar intensive quantity $q$. While multicomponent alloys have been covered
since the original introduction of the cluster expansion \cite{sanchez:cexp}%
, handling multisublattice systems (in which different sites in the unit
cell can host different binary sets of atoms) is a more recent extension 
\cite{tepesch:coupled}. ATAT implements a superset of these two formalisms,
allowing different sites to host an arbitrary (and potentially different)
number of elements, some of which may be common across sublattices. These
sublattices may or may not consist of sets of symmetrically equivalent
sites.\ To fix the ideas, refer to Table \ref{tablat} for an example
exploiting all of these extensions. The multicomponent multisublattice
features of ATAT have been used, for instance, in \cite%
{avdw:smceo2,ghosh:AlZnTi,shin:ternsqs}.

\begin{table}%

\caption{\label{tablat} Example of a lattice, based on the Martensite system,
written in ATAT's "lattice input file" format.}%

\begin{tabular}{ll}
\texttt{1 1 1.05 90 90 90} & (Tetragonal coordinate system in $a$ $b$ $c$ $%
\alpha $ $\beta $ $\gamma $ notation) \\ 
\texttt{0.5 \ 0.5 \ 0.5} & (Primitive unit cell: one vector per line \\ 
\texttt{0.5 -0.5 \ 0.5} & expressed in multiples of the above coordinate \\ 
\texttt{0.5 \ 0.5 -0.5} & system vectors) \\ 
\texttt{0 \ \ 0 \ \ 0 \ \ Fe,Ni,Cr} & (Ternary sublattice) \\ 
\texttt{0 \ \ 0.5 0.5 C,Vac} & (Binary sublattice, where Vac stands for
\textquotedblleft vacancy\textquotedblright , \\ 
\texttt{0.5 0 \ \ 0.5 C,Vac} & with a second symmetrically equivalent site
\\ 
\texttt{0.5\ 0.5 0 \ \ C,Vac} & and a third inequivalent site due to the
tetragonality)%
\end{tabular}

\end{table}%

The remainder of this section introduces the requisite extensions to the
cluster expansion formalism. The reader is referred to earlier reviews on
the topic \cite%
{sanchez:cexp,ducastelle:book,fontaine:clusapp,zunger:NATO,zunger:scord,wolverton:srorev,ceder:oxides,asta:fppheq}
or to the first part of this manual \cite{avdw:atat} for a gentler
introduction to the topic.

A \emph{configuration} $\sigma $ is represented by a vector of occupation
variables $\sigma _{i}$ indicating which type of atom sits on lattice site $%
i $. In an alloy where site $i$ can host $M_{i}$ components,\ $\sigma _{i}$
could take any value from $0$ to $M_{i}-1$. This type of representation of
an alloy is also known as a \emph{lattice gas model}. The CE takes the
general form 
\begin{equation}
q\left( \sigma \right) =\sum\nolimits_{\alpha }m_{\alpha }J_{\alpha
}\left\langle \Gamma _{\alpha ^{\prime }}\left( \sigma \right) \right\rangle
_{\alpha }
\end{equation}%
where we have used the following definitions:

\begin{itemize}
\item \vspace*{-0.05in}$\alpha $ is a \emph{cluster}. In the simple case of
a binary alloy, a cluster can be described by a vector of elements $\alpha
_{i}$ equal to one or zero depending whether site $i$ belongs to the
cluster\ or not. If site $i$ can host $M_{i}$ components,\ $\alpha _{i}$ can
take any values from $0$ to $M_{i}-1$, with $0$ indicating that site $i$
does not belong to cluster $\alpha $, while the other positive values
reflect various possible functional form dependence of the energy of cluster 
$\alpha $ on the occupation $\sigma _{i}$ of site $i$.

\item \vspace*{-0.05in}The sum is over all possible clusters $\alpha $ that
are mutually symmetrically distinct under the space group of the underlying
lattice.\footnote{%
Hence, out of each set of equivalent clusters, one representative cluster is
kept.} Note that the determination of the space group of the lattice must
account for the fact that sites hosting different sets of atoms are
considered symmetrically distinct.

\item \vspace*{-0.05in}The average $\left\langle \cdots \right\rangle
_{\alpha }$ is over all clusters $\alpha ^{\prime }$ that are equivalent by
symmetry to cluster $\alpha $.

\item \vspace*{-0.05in}$\Gamma _{\alpha ^{\prime }}\left( \sigma \right) $
are \emph{cluster functions}. They are selected to be of the form%
\begin{equation}
\Gamma _{\alpha }\left( \sigma \right) =\prod\nolimits_{i}\gamma _{\alpha
_{i},M_{i}}\left( \sigma _{i}\right)  \label{eqgammaprod}
\end{equation}%
where $\gamma _{\alpha _{i},M_{i}}\left( \sigma _{i}\right) $ satisfies $%
\gamma _{0,M_{i}}\left( \sigma _{i}\right) =1$ and the following
orthogonality condition%
\begin{equation}
\frac{1}{M_{i}}\sum_{\sigma _{i}=0}^{M_{i}-1}\gamma _{\alpha
_{i},M_{i}}\left( \sigma _{i}\right) \gamma _{\beta _{i},M_{i}}\left( \sigma
_{i}\right) =\left\{ 
\begin{array}{l}
1\text{ if }\alpha _{i}=\beta _{i} \\ 
0\text{ otherwise}%
\end{array}%
\right. .  \label{assortho}
\end{equation}%
(In binary alloys, a common choice is $\gamma _{0,2}\left( 0\right)
=1,\gamma _{0,2}\left( 1\right) =1,\gamma _{1,2}\left( 0\right) =-1,\gamma
_{1,2}\left( 1\right) =+1$.) Although the product (\ref{eqgammaprod}) is, in
principle, over all lattice sites, the choice $\gamma _{0,M_{i}}\left(
\sigma _{i}\right) =1$ ensures that it reduces to a product over sites
within cluster $\alpha $ only.

\item \vspace*{-0.05in}$m_{\alpha }$ are multiplicities indicating the
number of clusters (e.g. per unit cell) equivalent to $\alpha $ by symmetry.

\item \vspace*{-0.05in}$J_{\alpha }$ are expansion coefficients to be
determined. They are also called Effective Cluster Interactions (ECI).
\end{itemize}

It can be shown that when \emph{all} clusters $\alpha $ are considered in
the sum, the cluster expansion is able to represent \emph{any} function $%
q\left( \sigma \right) $ of configuration $\sigma $ by an appropriate
selection of the values of $J_{\alpha }$. The crux of the argument leading
to that conclusion lies in establishing that the $\Gamma _{\alpha }\left(
\sigma \right) $ form an orthogonal basis for the space of functions of
configurations according to a suitably defined inner product. In the general
case considered here, the original proof \cite{sanchez:cexp} needs to be
extended as follows. Given two cluster functions $\Gamma _{\alpha }$ and $%
\Gamma _{\beta }$ associated with two corresponding clusters $\alpha $ and $%
\beta $, define the inner product:%
\begin{equation*}
\left\langle \Gamma _{\alpha },\Gamma _{\beta }\right\rangle =\sum_{\sigma
}\Gamma _{\alpha }\left( \sigma \right) \Gamma _{\beta }\left( \sigma \right)
\end{equation*}%
where the sum is over all possible configurations $\sigma $ (i.e., each $%
\sigma _{i}$ ranges from $0$ to $M_{i}-1$ in lexicographic order). We can
then note that%
\begin{eqnarray*}
\left\langle \Gamma _{\alpha },\Gamma _{\beta }\right\rangle &=&\frac{1}{%
\prod_{i}M_{i}}\sum_{\sigma }\left( \prod\nolimits_{i}\gamma _{\alpha
_{i},M_{i}}\left( \sigma _{i}\right) \right) \left( \prod\nolimits_{i}\gamma
_{\beta _{i},M_{i}}\left( \sigma _{i}\right) \right) \\
&=&\frac{1}{M_{1}}\sum_{\sigma _{1}=0}^{M_{1}-1}\frac{1}{M_{2}}\sum_{\sigma
_{2}=0}^{M_{2}-1}\ldots \prod\nolimits_{i}\left( \gamma _{\alpha
_{i},M_{i}}\left( \sigma _{i}\right) \gamma _{\beta _{i},M_{i}}\left( \sigma
_{i}\right) \right) \\
&=&\prod\nolimits_{i}\left[ \frac{1}{M_{i}}\sum_{\sigma
_{i}=0}^{M_{i}-1}\left( \gamma _{\alpha _{i},M_{i}}\left( \sigma _{i}\right)
\gamma _{\beta _{i},M_{i}}\left( \sigma _{i}\right) \right) \right]
\end{eqnarray*}%
where the term in bracket is zero for any $i$ such that $\alpha
_{i}\not=\beta _{i}$ by Equation (\ref{assortho}). If no such $i$ exist,
then each term in bracket is equal to $1$. It follows that $\left\langle
\Gamma _{\alpha },\Gamma _{\beta }\right\rangle =0$ if the clusters $\alpha $
and $\beta $ differ (even by a single site) and is equal to $1$ if they are
identical. This establishes that the $\Gamma _{\alpha }$ are orthogonal.
Moreover, there are as many different $\alpha $ as there are different $%
\sigma $ (as $\alpha _{i}$ and $\sigma _{i}$ both can take on $M_{i}$
distinct values), so the \textquotedblleft number\textquotedblright\ of $%
\Gamma _{\alpha }$ equals\footnote{%
This statement can be made more rigorous by considering a sequence of
finite system (with periodic boundary conditions) whose size increases to
infinity.} the \textquotedblleft dimension\textquotedblright\ of the
configurational space $\sigma $. It follows that the $\Gamma _{\alpha }$
form a complete orthogonal basis.
This result is not obvious when lattice sites may host a different number
of species (i.e. when $M_i$ is not constant). In that case, the cluster functions
(Equation (\ref{eqgammaprod})) may involve mixed products between entirely different types
of $\gamma _{\alpha _{i},M_{i}}\left(\sigma _{i}\right) $.

While, the above discussion is valid for any $\gamma _{\alpha
_{i},M_{i}}\left( \sigma _{i}\right) $ satisfying (\ref{assortho}), a
specific $\gamma _{\alpha _{i},M_{i}}\left( \sigma _{i}\right) $ must be
selected in practice. By default, the ATAT uses the following choice:%
\begin{equation*}
\gamma _{\alpha _{i},M_{i}}\left( \sigma _{i}\right) =\left\{ 
\begin{array}{cl}
1 & \text{if }\alpha _{i}=0 \\ 
-\cos (2\pi \left\lceil \frac{\alpha _{i}}{2}\right\rceil \frac{\sigma _{i}}{%
M}) & \text{if }\alpha _{i}>0\text{ and odd} \\ 
-\sin (2\pi \left\lceil \frac{\alpha _{i}}{2}\right\rceil \frac{\sigma _{i}}{%
M}) & \text{if }\alpha _{i}>0\text{ and even}%
\end{array}%
\right.
\end{equation*}%
where $\left\lceil \ldots \right\rceil $ denotes the \textquotedblleft round
up\textquotedblright\ operation and where both $\alpha _{i}$ and $\sigma
_{i} $ can range from $0$ to $M_{i}-1$. Note that these functions reduce to
the single function $-\left( -1\right) ^{\sigma _{i}}$ in the binary case ($%
M_{i}=2$). User-specified $\gamma _{\alpha _{i},M_{i}}\left( \sigma
_{i}\right) $ can be provided by (i) editing and following the instructions
given in the \texttt{corrskel.c++} file and by (ii) specifying the \texttt{%
-crf=[keyword]} option on the command line of ATAT commands, where \texttt{%
[keyword]} is a unique identifier.

Another nonobvious aspect of the above cluster expansion generalization is
that the relationship between the point correlations and the composition
variables is no longer as trivial as in the well-known single sublattice
binary case. In general, there are three different composition-type
quantities to consider, each having a specific use.

\begin{enumerate}
\item The point correlation vector $\rho $ contains all the expected values
of the cluster functions $\left\langle \Gamma _{\alpha }\left( \sigma
\right) \right\rangle $ associated with a cluster $\alpha $ such that $%
\alpha _{i}$ is nonzero for a single site $i$. They enter the expression of
the cluster expansion which is used to predict energies either during the
ground state search procedure or in the Monte Carlo simulations.

\item The \textquotedblleft nonredundant\textquotedblright\ concentration
vector $c$ are those that remain after all linearly dependent concentrations
have been eliminated (of course, there are multiple valid choices of
nonredundant\ concentrations --- one convention must be arbitrarily
selected). These linear dependencies arise from the constraint that the
compositions on each sublattice sum up to one. The nonredundant
concentrations are used for the convex hull construction in the ground state
search procedure.

\item The full vector $x$ of concentrations has no algorithmic use but
provides the most user-friendly output.
\end{enumerate}

These quantities are linearly related:%
\begin{eqnarray}
c &=&C\rho +c_{0}  \label{eqgivec} \\
x &=&X\rho +x_{0}  \label{eqgivex}
\end{eqnarray}%
where $C$ and $X$ are constant matrices while $c_{0}$ and $x_{0}$ are
constant vectors. In general, the matrices $C$ and $X$ are rectangular and
not invertible. (These matrices can be output within ATAT with the \texttt{%
corrdump -pcm} command, if the user provides a lattice input file \texttt{%
lat.in}.)\ Appendix \ref{appconv} describes how these matrices are
calculated.

The matrix $X$ plays a special role in the conversion of chemical potentials 
$\mu $ into shifts of the effective cluster interactions $J_{\alpha }$, as
would be needed in a grand canonical Monte Carlo simulation. Note that the
chemical potential contribution to any of the grand canonical thermodynamic
functions can be written as%
\begin{equation*}
\mu ^{T}x=\mu ^{T}X\rho +\mu ^{T}x_{0}=\left( X^{T}\mu \right) ^{T}\rho +\mu
^{T}x_{0}=\sum_{j}\left( X^{T}\mu \right) _{j}\rho _{j}+\mu ^{T}x_{0}
\end{equation*}%
from which it can be inferred that $\mu ^{T}x_{0}$ is the shift induced in
the empty cluster ECI, while the $j$-element of the vector $X^{T}\mu $ is
the shift induced in the ECI associated with point correlation $\rho _{j}$.

\subsection{Multicomponent cluster expansion construction}

The \texttt{mmaps} command is the multicomponent version of the \texttt{maps}
command (discussed in \cite{avdw:atat,avdw:maps}) and works in a similar
fashion. It gradually constructs an increasingly more accurate cluster
expansion by repeatedly invoking a first-principles code through auxiliary
scripts (see Sections \ref{secnonconf}, \ref{secabi} and the documentation
of the \texttt{pollmach} and \texttt{runstruct\_xxxx} commands for details).
The main difference between the \texttt{mmaps} and \texttt{maps} codes is
that, given the larger number of components allowed in \texttt{mmaps}, the
input and output files have to contain more information. The main input file
(typically called \texttt{lat.in}), which describes the lattice, is a
natural extension of the \texttt{maps} input file, as shown in Table \ref%
{tablat}.

A second, optional, input file (typically called \texttt{crange.in})
provides the ability to specify a region to explore in composition space.
This is useful in systems where the provided lattice would be mechanically
or thermodynamically unstable beyond certain composition limits. For
instance, given the lattice shown in Table \ref{tablat}, one could specify:%
\newline
\hspace*{0.5in}%
\texttt{1*Ni+1*Cr{<}=0.2}\newline
\hspace*{0.5in}%
\texttt{1*C{<}=0.1}\newline
where each atom name stands for its overall concentration (it is not the
concentration on a sublattice). Only linear constraints are supported (and
are implicitly combined with a logical \textquotedblleft
and\textquotedblright , so that the allowed regions are necessarily convex).

This file not only controls the range of concentration of the structures
generated by \texttt{mmaps}, but also the composition range where the
correct ground states are enforced in the fitting process. Note that,
occasionally, structures outside the specified range may be generated in
order to ensure that the ground states found within the specified
composition range are not masked by ground states outside of it (\texttt{%
mmaps} internally generates structures at all compositions in order to check
for this).

Suitable composition limits are not always known in advance. When these
limits arise due to mechanical instabilities, one would notice that many of
the structures generated by \texttt{mmaps} relax to very different types of
lattices. A symptom of this is the inability to converge the cluster
expansion. A way to check for this problem is to use the \texttt{checkrelax}
utility, which calculates the mean square of the relaxation strain tensor
elements for each of the structures generated by \texttt{mmaps}. Note that,
by design, this command ignores the isotropic component of the strain. It
compares relaxed geometries in the \texttt{*/str\_relax.out} files to the
corresponding unrelaxed \texttt{*/str.out} files. Any value above \texttt{0.1%
} should be regarded as suspect. While having a few isolated overrelaxed
structures is not typically a problem, if virtually all structures beyond a
certain composition limit are relaxing excessively, then it is advisable to
specify a suitable \texttt{crange.in} file and restart \texttt{mmaps}. It is
also recommended to \textquotedblleft hide\textquotedblright\ any
overrelaxed structures by issuing the command\newline
\hspace*{0.5in}%
\texttt{touch [structure]/error}\newline
for each \texttt{[structure]} exhibiting overrelaxation (typically \texttt{%
[structure]} is a number referring to one of the subdirectories generated by 
\texttt{mmaps} and containing a structure's geometry). This command creates
files called \texttt{error} in the specified directory that \texttt{mmaps}
will perceive as a signal to ignore the structure in question.

The relevant composition region is sometimes limited by thermodynamic
instabilities. This would be detected, for instance, by doing a separate
cluster expansion on a different lattice or by computing the formation
energies of known compounds exhibiting a different lattice. In such cases,
however, it is not essential to limit the compositions range, since the
cluster expansion formalism remains valid for metastable alloys. However,
unless the metastable region is of direct interest, a more rapidly
converging cluster expansion may be obtained by specifying a suitable 
\texttt{crange.in} file and by excluding non-ground state structures in the
metastable region with the \texttt{touch} command, as explained above.

It is sometimes tempting to perform a cluster expansion under \emph{equality}
--- rather than inequality --- composition constraints. This is relevant for
ionic systems, where it may be most natural to only keep structures that are
\textquotedblleft charge balanced\textquotedblright\ under the assumption
that all species keep their nominal charges. This is not implemented in 
\texttt{mmaps}, because it is very risky approach. A cluster expansion
providing an excellent\ fit to charge-balanced structures may inadvertently
predict non-charge-balanced ground states. A way to prevent such problems 
\cite{tepesch:coupled}\ is to restrict compositions to lie in a
nonzero-width slice around the charge-balanced hyperplane, as in the
following example for the Samarium-doped Ceria system \cite{avdw:smceo2}:%
\newline
\hspace*{0.5in}%
\texttt{-2*O+4*Ce+3*Sm{<}=0.2}\newline
\hspace*{0.5in}%
\texttt{-2*O+4*Ce+3*Sm{>}=-0.2}\newline
\texttt{mmaps} will, if necessary, sample structures outside of the
specified range to ensure that no spurious non-charge-balanced ground states
exist.

Compared to \texttt{maps}, the \texttt{mmaps} code has a number of new
output files. The \texttt{atoms.out} file lists all atomic species given in
the input files \emph{in the order that will be used to report composition
variables in all output files}. The \texttt{ref\_energy.out} file reports
the atomic reference energies used to calculate formation energies in all
output files (in the same order as in \texttt{atoms.out}). These reference
energies are the energies of the pure elements, if they are included in the
fit. Otherwise, if composition constraints are imposed,\footnote{%
or if some elements are excluded from some sublattices, thus effectively
restricting the composition range.} the structures with the most extreme
compositions\footnote{%
Specifically, those with the largest sum of squared concentrations.} are
used as reference. These non-standard references are converted into atomic
energies before being output in \texttt{ref\_energy.out}. This default
behavior can be overridden by providing user-specified reference energies in
a file called \texttt{ref\_energy.in}.

\begin{figure}
\centerline{\epsfbox{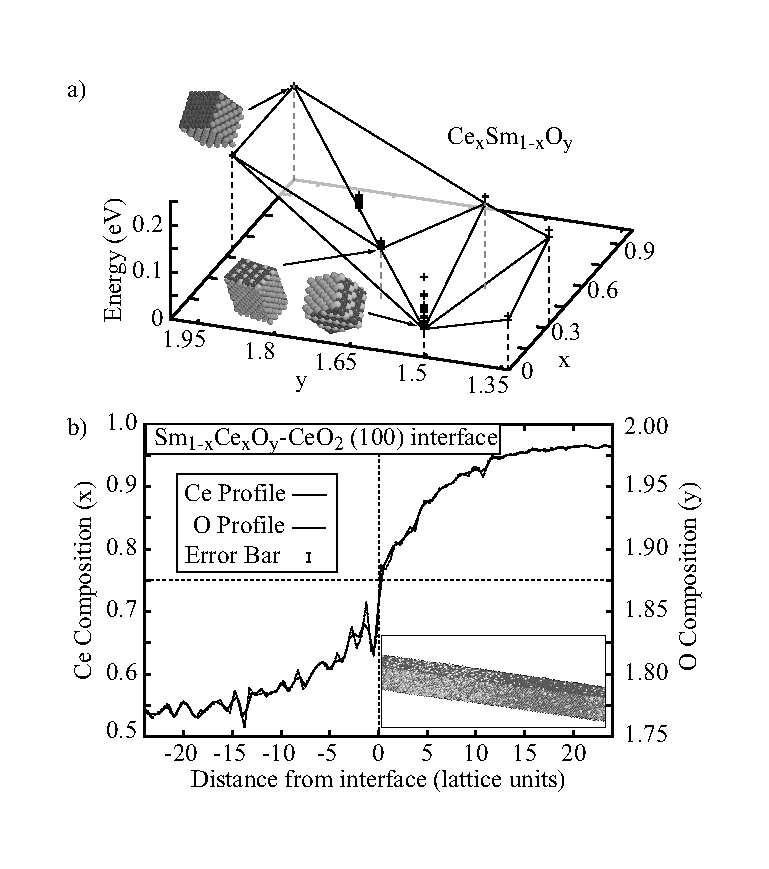}}
\caption{\label{figceria}
Computational thermodynamic study of the Samarium-doped Ceria system
\cite{avdw:smceo2}.
a) Example of ground state search via a convex hull construction
in energy-composition space.
b) Equilibrium composition profile across (100) interface obtained through a
hybrid canonical/grand-canonical multisublattice multicomponent
Monte-Carlo simulation.
Inset: Snapshot of the 70,000-atom simulation cell used.
}
\end{figure}

While the ground state compositions and energies are output in \texttt{gs.out%
}, the file \texttt{gs\_connect.out} provides the additional information
needed, in multicomponent systems, to describe how these ground states are
connected by tie lines. Each facet of the ground state convex hull
corresponds to a line in the \texttt{gs\_connect.out} file listing the
structure names associated with the vertices of that facet. An example of a
convex hull in energy-composition space constructed with ATAT is shown in
Figure \ref{figceria}a).

Setting up grand-canonical Monte Carlo simulations (as discussed in the next
section) demands the knowledge of the chemical potentials that stabilize
various phase equilibria of the system. Since this is not obvious to figure
out in a multicomponent system, \texttt{mmaps} generates a file \texttt{%
chempot.out} containing:

\begin{enumerate}
\item The values of the chemical potentials that stabilize each
fixed-composition multiphase equilibria of the system at absolute zero\
(e.g. in a $n$-nary system, all the $n$-phase equilibria). Each vector of
chemical potential correspond to one facet of the ground state convex hull.

\item For each ground state, a value of the chemical potentials that
stabilize it at absolute zero. This is not a uniquely defined quantity ---
many possible chemical potentials values are able to stabilize a given
ground state. Among the many possible values, \texttt{mmaps} calculates one
by simply averaging the chemical potentials associated with all facets that
intersect a given ground state.
\end{enumerate}

The remaining output files are the same as in \texttt{maps}, except that
multiple concentration variables are output whenever necessary: \texttt{%
maps.log}, \texttt{fit.out}, \texttt{predstr.out}, \texttt{gs.out}, \texttt{%
gs\_str.out}, \texttt{eci.out}, \texttt{clusters.out}.

\subsection{Hybrid Multicomponent Monte Carlo simulations}

The \texttt{memc2} code implements general hybrid canonical/grand-canonical
lattice gas Monte Carlo simulations in multicomponent systems. As most of
the features are similar to the previous binary version, \texttt{emc2},
described in \cite{avdw:atat,avdw:emc2}, we focus here on the differences.

To function, this code requires at least 4 input files: \texttt{lat.in}
(identical to the one input into \texttt{mmaps}), \texttt{clusters.out}, 
\texttt{eci.out}, \texttt{gs\_str.out}, where the latter three are generated
by \texttt{mmaps}. Given the larger number of composition variables, the
range of temperatures and chemical potentials scanned in the course of
grand-canonical run are no longer specified on the command-line. Instead,
the \texttt{control.in} input file must be provided in the following format.
The first line specify the initial conditions and has the form (for an $n$%
-component system):\newline
\hspace*{0.5in}%
\texttt{[}$T$\texttt{] [}$\mu _{1}$\texttt{] \ldots\ [}$\mu _{n}$\texttt{]}%
\newline
where $T$ is temperature and $\mu _{1},\ldots ,\mu _{n}$ $\,$are the
chemical potentials for each chemical species, in the same order as in the 
\texttt{atoms.out} file generated by \texttt{mmaps}.\footnote{%
Alternatively, see the ouput of \texttt{corrdump -pcm}.} Each subsequent
line of this file indicates one of the axes along which to scan and has the
format:\newline
\hspace*{0.5in}%
\texttt{[}$T$\texttt{] [}$\mu _{1}$\texttt{] \ldots\ [}$\mu _{n}$\texttt{] [}%
$s$\texttt{]}\newline
where $s$ is the the number of steps made between the initial conditions and
the final conditions given on the line. Table \ref{tabctrl} gives an example
of such file.

\begin{table}%

\caption{\label{tabctrl}
Example of {\tt control.in} file indicating a scan of the region in ($T,\mu_1,\mu_2,\mu_3$)-space
defined by $100 \leq T < 200$ and $0 \leq \mu_1 < 1.0$ and $0 \leq \mu_2 < 0.5$
and $\mu_3 = 0.0$ with a $10 \times 5 \times 5$ grid.
}

\texttt{100 \ 0.0 0.0 0.0}

\texttt{200 \ 0.0 0.0 \ 0.0 \ 10}

\texttt{100 \ 0.0 0.5 \ 0.0 \ 5}

\texttt{100 \ 1.0 0.0 \ 0.0 \ 5}

\end{table}%

A few important notes:

\begin{enumerate}
\item Temperatures are given in units of energy unless the \texttt{-k} or 
\texttt{-keV} options are used (the latter indicates temperature in K if
energies are in eV).

\item By default, the finals conditions are excluded from the scan (in the
example above the values of the first chemical potential scanned (see the
last line) are 0.0\ 0.2 0.4 0.6 0.8). This behavior can be changed with the 
\texttt{-il} option, to give: 0.0 0.25 0.5 0.75 1.0). Alternatively, the 
\texttt{-hf} option gives: 0.1 0.3 0.5 0.7 0.9.

\item The \texttt{mmaps} code generates a file called \texttt{chempot.out}
which contains special values of the chemical potential that stabilize
various types of equilibria. These values are useful guidelines to select
relevant regions in $\mu $-space to scan.
\end{enumerate}

Canonical or hybrid canonical/grand-canonical simulations can be carried out
by specifying, in a file called \texttt{conccons.in}, a set of linear
constraints on the composition that must hold throughout the simulation. The
code automatically generates a complete list of allowed multi-site Monte
Carlo moves. It does so by first considering all possible single-atom
identity flips ($k=1$), then all possible two-atom identity flips ($k=2$),
etc. Any move violating the composition constraints is eliminated. After all 
$k$-atom flips that satisfy the constraints have been considered, the code
calculates the linear subspace spanned by the composition changes associated
these flips. The process stops if this linear subspace has the same
dimension as the space of allowed compositions. The moves generated by this
process are guaranteed to satisfy detailed balance because for every move
kept, its reverse move is also kept.

The format of the \texttt{conccons.in} file is similar to the \texttt{%
crange.in} file of \texttt{mmaps}. For instance:\newline
\texttt{-2*O+4*Ce+3*Sm=0}\newline
would ensure overall charge balance in the fluorite-type Ce$_{x}$Sm$_{1-x}$O$%
_{y}$Vac$_{2-y}$ system. Constraining composition variables is particularly
useful to pin down the location of an interface in a multiphase interfacial
thermodynamic calculation, because the phase fractions are uniquely
determined by the overall composition.\ An example of this use is
illustrated in Figure \ref{figceria}b).

Any number of constraints can be given up to the point where all
concentration variables are fixed, which results in a fully canonical
simulation. Note that constant composition simulations do not evade the need
to make sure that the cluster expansion does not exhibit spurious ground
states away from the specified hyperplane. If such ground states exist, the
Monte Carlo code will reveal a phase separation into regions that do not
locally satisfy the composition constraint, although the overall simulation
cell does. Note that the constant specified on the right-hand side of the 
\texttt{conccons.in} file is ignored --- this file constrains changes in
composition only. The user-supplied initial atomic configuration (via the 
\texttt{-g=[ground state number]} or \texttt{-is=[file]} options) implicitly
determines the right-hand side constants.

The \texttt{memc2} code outputs all of thermodynamic quantities it computes
on the standard output and in a log file \texttt{mc.out}. Given the large
(and variable) number of variables in a multicomponent system, the code
writes a \texttt{mcheader.out} file indicating the content of each column of
the main output file. By default all output quantities are grand-canonical
in nature (i.e. they contain a \textquotedblleft $-\mu \cdot x$%
\textquotedblright\ term, where $x$ is composition and $\mu $ chemical
potential), regardless of the presence of composition constraints. To obtain
\textquotedblleft canonical\textquotedblright\ rather than grand-canonical
quantities use the \texttt{-g2c} option.

In addition to performing Monte Carlo simulations, the \texttt{memc2} code
reports thermodynamic quantities calculated via the low-temperature
expansion (indicated by a \texttt{\_lte} suffix ) or the mean-field
approximation\footnote{%
While the \texttt{emc2} code also reports the high-temperature limit, this
was abandoned in the \texttt{memc2} code since it is identical the
mean-field values for the pure phases.} (indicated by a \texttt{\_mf} suffix
).\ A useful feature is the ability to skip Monte Carlo simulations whenever
the free energies obtained via mean-field approximation and the low
temperature expansion fall within a specified threshold (using the \texttt{%
-mft=[tolerance]} option), as this indicates that either one is sufficiently
accurate. These approximations are also used to set initial values of the
free energy when performing thermodynamic integrations. By default, if the
initial configuration is a ground state (\texttt{-gs=[a positive integer]})\
the low-temperature expansion provides the starting point while Monte Carlo
averages are used to update the free energy as temperature and chemical
potentials are varied. If the initial configuration is the fully disordered
state (\texttt{-gs=-1}) a mean-field approximation (i.e. high-temperature
expansion) is used as the starting point instead.

Finally, it is possible to specify temperature-dependent interactions (in a
file called \texttt{teci.out}, which, when present, takes precedence over 
\texttt{eci.out}). This feature can be used to account for
nonconfigurational sources of entropy through a coarse-graining scheme \cite%
{avdw:vibrev,ceder:ising}. 
Introducing temperature-dependent interactions is not completely trivial,
as various averages (such as the
internal energy $E$) must correctly account for the temperature dependence
of the interactions in order for the following thermodynamic relationship
(upon which thermodynamic integration is based) to hold: 
\begin{equation*}
\frac{\partial \left( \beta F\right) }{\partial \beta }=E
\end{equation*}%
where $\beta =\left( k_{B}T\right) ^{-1}$ and $F$ is the Helmholtz free
energy. To illustrate this, we express the free energy $F$ in terms of the
partition function and the coarse grained free energy\footnote{%
In a coarse-graining framework, each microscopic configurational state $i$ has a
free energy that can be expressed in terms of a partition function summing
over all vibrational and electronic excitations $j$ associated with a given
configuration $i$ on the lattice: $F_{i}=-\beta ^{-1}\ln \left(
\sum_{j}e^{-\beta E_{j}}\right) $.} $F_{i}$ of each microscopic state $i$:%
\begin{equation*}
\frac{\partial \left( \beta F\right) }{\partial \beta }=-\frac{\partial }{%
\partial \beta }\ln \left( \sum_{i}e^{-\beta F_{i}}\right) =\frac{\sum_{i}%
\left[ F_{i}+\beta \left( \partial F_{i}/\partial \beta \right) \right]
e^{-\beta F_{i}}}{\sum_{i}e^{-\beta F_{i}}}.
\end{equation*}%
Hence, the quantity to average over the Monte Carlo steps is $\left[
F_{i}+\beta \left( \partial F_{i}/\partial \beta \right) \right] $ instead
of just $F_{i}$, as a naive analogy with the constant interaction case would
have suggested.

The next section describes how to proceed to obtain suitable
temperature-dependent interactions that model vibrational and electronic
excitations.

\section{Nonconfigurational sources of entropy}

\label{secnonconf}

\subsection{Lattice vibrations}

The ATAT software offers two techniques to calculate vibrational free
energies.

The \texttt{fitfc} code constructs a Born-von K\'{a}rm\'{a}n spring model by
fitting the reaction forces resulting from imposed atomic displacements in a
supercell calculation \cite{avdw:vibrev,wei:vib,gdg:vib2}. This is the
preferred method\footnote{%
In principle, linear response calculations \cite%
{baroni:lr,giannozzi:lr,gonze:lr,gonze:lr_rev1,gonze:lr_rev2,gironcoli:lr}
are even more preferable in the sense that they account for infinite range
interactions, which is important in the case of ionic materials. Linear
response calculations are to be performed within the ab initio code itself
and ATAT consequently does not implement them.} for very accurate
calculations, since the range of interactions of the model's effective
springs can be increased until the desired accuracy has been reached.
However, it is computationally demanding to calculate vibrational free
energies for a large number of structures in this fashion (as would be
required for cluster expansion construction).

For this reason, ATAT includes an alternative method: The
stiffness-vs.-length scheme \cite{avdw:vibrev}, which is implemented in\ the 
\texttt{fitsvsl} and \texttt{svsl} codes. This approach exploits the fact
that the effective springs of a Born-von K\'{a}rm\'{a}n model are often
transferable from one structure to another (after controlling for simple
factors such as\ bond length \cite%
{avdw:pd3v,morgan:aruba,wu:svsl,avdw:vibrev}, or composition \cite{liu:cdtfc}%
), thus reducing the number of spring constant calculations needed. This
method's accuracy is limited by the transferability assumption itself, which
typically does not extend beyond nearest neighbor springs. Fortunately, this
assumption is verifiable and we shall see how.

\subsubsection{The fitfc code}

\label{secfitfc}Calculations with the fitfc code proceed in a series of
steps:

\begin{enumerate}
\item The structure of interest first needs to be fully relaxed. The code
expects the relaxed geometry in a file called \texttt{str\_relax.out}. This
file's format is common to all ATAT tools and is described in \cite%
{avdw:atat} as well as in the help obtained via \texttt{mmaps -h}.\ The 
\texttt{fitfc} command also expects a \texttt{str.out} file containing the 
\textbf{unrelaxed} geometry (which may be the same as the relaxed geometry,
if desired). The unrelaxed geometry will be used to determine the neighbor
shells and measure distances between atoms. The rationale for allowing for
two types of input structures is that it is easier for most users to specify
or identify shells\ of neighbor using \textquotedblleft
idealized\textquotedblright\ structures (e.g. hcp with ideal $c/a$ ratio)
rather than using the fully relaxed structures. Moreover, in the context of
cluster expansion construction, where phonon calculations on multiple
structures are needed, the neighbor shell distances are directly
transferable between unrelaxed structures but not between relaxed structures.%
\newline
Typically the user would specify the \texttt{str.out} and then obtain the 
\texttt{str\_relax.out} file by running an ab initio code with a command of
such as \texttt{runstruct\_xxxx}, where \texttt{xxxx} stands for the name of
the ab initio code used (see Section \ref{secabi} below for calling ab
initio codes within ATAT). Of course, the input files for the \texttt{xxxx}
code must indicate that all degrees of freedom need to be relaxed.
Alternatively, file \texttt{str.out} and \texttt{str\_relax.out} files could
have been automatically generated in a prior \texttt{mmaps} execution (in
fact, the perhaps odd extension \texttt{.out} reflects the fact that these
files are typically outputs of the \texttt{mmaps} and \texttt{runstruct\_xxxx}
codes.).

\item Next, a set of perturbed geometries must be generated. A typical
command line is as follows:\newline
\hspace*{0.5in}%
\texttt{fitfc -er=11.5 -ns=3 -ms=0.02 -dr=0.1}\newline
The \texttt{-er} option specifies how far apart the periodic images of the
displaced atom must be (as measured in the unrelaxed structure \texttt{%
str.out} and in the same units, usually \AA ). The code uses this
information to finds the smallest supercell satisfying this constraint.
While this is the only required parameter, the user has control over other
parameters.\newline
\texttt{-dr}\ specifies the magnitude of the displacement (default:\ 0.2 \AA %
) of the perturbed atom (in the same units as in the \texttt{str.out} file,
typically \AA )\newline
\texttt{-ns} specifies the number of different isotropic strain levels at
which phonon calculations will be performed. Specifying \texttt{-ns=1}
implies a purely harmonic model (the default),\ while values greater than 1
will invoke a quasiharmonic model. The latter accounts for thermal expansion
by allowing the phonon frequencies to be volume-dependent. In subsequent
steps, the code will determine the volume as a function of temperature by
minimizing the free energy.\newline
\texttt{-ms} specifies the maximum level of strain (e.g. 0.02 signifies a
2\% elongation along every direction).\newline
The above command writes out a series of subdirectories of the general form 
\texttt{vol\_*}, one for each level of strain (\texttt{*} refers to the
usual UNIX\ wildcard).\newline
If the structure has cubic symmetry and no internal degrees of freedom 
\textbf{or} if one only wishes to use the harmonic approximation, the fitfc
command should be invoked with the \texttt{-nrr} option (do Not ReRelax).\
In these case,\ the relaxation step \ref{steprr} below can be skipped.

\item \label{steprr}Each \texttt{vol\_*} subdirectory now contains a \texttt{%
str.out} file which is stretched version of the main \texttt{str\_relax.out}
file provided. (This naming convention is a bit confusing but is dictated by
the fact that these \texttt{vol\_*/str.out} files are the starting point of
another relaxation step.) The ab initio code then needs to be invoked to
rerelax the geometry at the various levels of imposed strain and obtain the
energy as a function of strain. Typically, this is achieved by typing:%
\footnote{%
The \texttt{pollmach} command looks for files called \texttt{wait} in all
subdirectories as a signal that the specified command must be run the
directory containing that file. Hence it is prudent to double check that
there are no leftover \texttt{wait} files from earlier calculations: \texttt{%
find . -name wait}}\newline
\hspace*{0.5in}%
\texttt{pollmach -e runstruct\_xxxx -w\ ionrelax.wrap}\newline
The \texttt{pollmach -e} \ldots\ prefix causes the command specified next to
be run in all subdirectories containing a file called \texttt{wait} (a file
automatically written by \texttt{fitfc} with this purpose in mind) before
exiting. Here, the command to be run is \texttt{runstruct\_xxxx}, which
invokes the ab initio code \texttt{xxxx}. The option \texttt{-w\
ionrelax.wrap} specifies an alternative user-specified input file for the ab
initio code, which must indicate that all degrees of freedom \textbf{except
volume} must be allowed to relax. This additional relaxation step is needed\
because stretching may have modified the equilibrium atomic positions. The
relaxed (but stretched) structures now reside in the \texttt{%
vol\_*/str\_relax.out} files while the corresponding energies will be in 
\texttt{vol\_*/energy}.

\item \label{fitfcstepfiles}Now, one needs to invoke \texttt{fitfc} again to
generate perturbations of the atomic position for each level of strain, e.g.:%
\newline
\hspace*{0.5in}%
\texttt{fitfc -er=11.5 -ns=3 -ms=0.02 -dr=0.1}\newline
This is exactly the same command as before but the code notices the presence
on the new files and proceeds further.\newline
At this point the files generated are arranged as follows. At the top level,
there is one subdirectory per level of strain (\texttt{vol\_*}, where * is
the strain in percent), and in each subdirectory, there are a number of
subsubdirectories, each containing a different perturbation. The
perturbation names have the form\newline
\hspace*{0.5in}%
\texttt{p[+/-][dr]\_[er]\_[index]},\newline
where \texttt{[pertmag]} is the number given by the \texttt{-dr} option, 
\texttt{[er]} the number given by the \texttt{-er} option and \texttt{[index]%
} is a number used to distinguish between different perturbations. Two
perturbations that differ only by their signs are sometimes generated and
are distinguished by a \texttt{+} or \texttt{-} prefix. (These opposite-sign
perturbations ensure that the effect of third-order force constants cancel
out exactly in the fit. Note that whenever the third-order terms cancel out
by symmetry, the code will recognize this and only generate the
\textquotedblleft positive\textquotedblright\ perturbation. If one is not
concerned with errors introduced by third-order force constants and wishes
to save time, \textquotedblleft negative\textquotedblright\ perturbations
can be ignored at this point by deleting the \texttt{vol\_*/p-*/wait}
files.) Each subdirectory of the form \texttt{vol\_*/p*} contains (i) the
ideal unrelaxed supercell in a \texttt{str\_ideal.out} file, (ii) the
relaxed but unperturbed supercell in a \texttt{str\_unpert.out} file and
(iii) the actual geometry of perturbed supercell calculation in a \texttt{%
str.out} file.

\item \label{fitfcstepvasp}The ab initio code must be invoked again, this
time to calculate reaction forces for each perturbation. This will typically
be accomplished by typing the now familiar construct:\newline
\hspace*{0.5in}%
\texttt{pollmach -e runstruct\_xxxx -w force.wrap }\newline
where the \texttt{force.wrap} must now indicate that \textbf{no} degrees of
freedom are allowed to relax.\footnote{%
A special note to VASP\ users:\ Smearing methods must be used for Brillouin
zone integration in force calculations, unless the system is insulating. Do
not use the DOSTATIC option in the \texttt{force.wrap} file (this is not
VASP keyword per se but it is interpreted by the \texttt{runstruct\_vasp}
command.).} Each subdirectory of the form \texttt{vol\_*/p*} will now
contain a \texttt{force.out} file (and a number of other ab initio
code-specific files not read by \texttt{fitfc}).\ A \texttt{str\_relax.out}
is also generated, although it contains the same geometry as the \texttt{%
str.out} file since no relaxation took place.\footnote{%
For some ab initio codes, the ordering of the atoms in \texttt{str.out} and 
\texttt{str\_relax.out} may differ, but \texttt{fitfc} is able to figure out
how to assign each force to the correct atom, provided the atoms in \texttt{%
str\_relax.out} and in \texttt{force.out} are in the same order. All the 
\texttt{runstruct\_xxxx} commands provided with ATAT ensure that this is the
case.} This concludes the \textquotedblleft generation\textquotedblright\
phase of the process.

\item Next, the code must fit the force constants and perform the actual
phonon calculations. This mode is invoked via the \texttt{-f} option:\newline
\hspace*{0.5in}%
\texttt{fitfc -f -fr=...}\newline
In addition, the range of the springs to be included in the fit is specified
using the mandatory parameter \texttt{-fr=...}. Usually, the range specified
with \texttt{-fr} should be no more than half the distance specified with
the \texttt{-er} option earlier. Distances are measured according to the
atomic positions given in \texttt{str.out} and in the same units (typically 
\AA ). It is a good idea to try different values of \texttt{-fr} (starting
with the nearest neighbor bond length) and check that the vibrational
properties converge as \texttt{-fr} is increased. The output files will be
described below.

\item Sometimes, the message '\texttt{Unstable modes found. Aborting.}' is
printed. This indicates that the structure considered \textbf{may} be
mechanically unstable. If, in addition, you see the warning '\texttt{%
Warning: p\ldots\ is an unstable mode}.', then the structure is certainly
mechanically unstable. Otherwise, it may be an artifact of the fitting
procedure. To find out, the code can generate perturbations along the
unstable directions and let the ab initio code calculate the reaction forces
which can then be included in the fit to settle this issue.\newline
First, to view the unstable modes, use the \texttt{-fu} option. The output
will be in vol\_*/unstable.out and has the form:\newline
\hspace*{0.5in}%
\texttt{u [index] [nb\_atom] [kpoint] [branch] [frequency]}\newline
\hspace*{0.5in}%
\texttt{[displacements...]}\newline
where \texttt{[index]} is a reference number, \texttt{[nb\_atom]} is the
size of the supercell needed to represent this mode and \texttt{%
[displacements]}is a vector of $3n$ elements defining the displacement mode,
where $n$ is the number of atoms in the unit cell. The other entries are
self-explanatory. (If this file contains only entries with
nb\_atom=too\_large, one needs to increase the \texttt{-mau} option beyond
its default of 64.) One of these modes can be selected to be written out to
disk with the option \texttt{-gu=[index]} where \texttt{[index]} is the one
reported in the \texttt{unstable.out} file. This new feature aimed at the
elimination of fictitious unstable modes was first used in \cite%
{benedek:acid,benedek:exch}.

\item The ab initio code should then be run in the subdirectory generated
(named \texttt{vol\_*/p\_uns\_[dr]\_[kmesh]\_[number]}) and rerun, from the
top-level directory,\ \texttt{fitfc -f -fr=...} \newline
If the message \textquotedblleft \texttt{Warning: p... is an unstable mode}%
\textquotedblright\ appears, then a true instability has been found. If only
\textquotedblleft \texttt{Unstable modes found. Aborting}\textquotedblright\
is printed, one may repeat the process until the message disappears or a
truly unstable mode is found.\newline
Note: The \texttt{-fn} option can be used if one wishes to generate a phonon
DOS even if there are unstable modes.\ The unstable modes will be shown as
negative frequencies. The \texttt{-fn} option eliminates the printed error
message but not the underlying problem.
\end{enumerate}

The output files are as follows:

\begin{enumerate}
\item \texttt{fitfc.log} : A general log file.

\item \texttt{vol\_*/vdos.out} : the phonon density of states for each
volume considered

\item \texttt{vol\_*/fc.out} : The force constants. For each force constant
a summary line, prefixed with \textquotedblleft svsl\textquotedblright\
gives: (i) the atomic species involved, (ii) the 'bond length', (iii) the
stretching and bending terms. This can be useful to check the validity of
the length-dependent transferable force constant approximation. Then, each
separate component of the force constant is given and, finally, their sum.

\item \texttt{vol\_*/svib\_ht} : gives the high-temperature limit of the
vibrational entropy (in units of Boltzman constant per atom) in the harmonic
approximation, excluding the configuration-independent contribution at each
unit cell volume considered (so, this just $-3$ times the average log phonon
frequency).

\item \texttt{fitfc.out} : gives along each row, the temperature, the free
energy, and the linear thermal expansion (e.g. 0.01 means that the lattice
has expansion by 1\% at that temperature).

\item \texttt{fvib} : gives only the free energy (this is the output file
that may be subsequently use to construct a cluster expansion)

\item \texttt{svib} : gives only the entropy (in energy/temperature, by
default, eV/K)

\item \texttt{eos0.out} : equation of state at 0K (each line reports one
level of strain and its associated energy)

\item \texttt{eos0.gnu} : gnuplot script to plot the equation of state
polynomial and data.
\end{enumerate}

The files within the subdirectories \texttt{vol\_*} are based on the
harmonic approximation for a given fixed volume while the files in the main
directory give the output of the quasiharmonic approximation if \texttt{ns}$%
>1$ and the harmonic approximation if \texttt{ns}$=1$.

Examples studies employing this code can be found in \cite%
{arroyave:znzr,benedek:acid,benedek:exch,avdw:culi}, among others.

\subsubsection{The fitsvsl and svsl codes}

The \texttt{fitsvsl} code determines bond stiffness vs bond length
relationship for the purpose of calculating vibrational properties (with the 
\texttt{svsl} code, described subsequently). Using this pair of codes pays
off only if vibrational properties of large number of similar structures are
needed, because the so-called transferable length-dependent force constants
can be determined from a small set of structures with the \texttt{fitsvsl}
code, while vibrational properties can then be quickly predicted for a much
larger set of structures with the \texttt{svsl} code.

\texttt{fitsvsl} requires the following files as an input.

\begin{enumerate}
\item A lattice file (\texttt{lat.in}) which allows the code to determine
what chemical bonds are expected to be present in the system. The format is
as described in the mmaps documentation (see \texttt{mmaps -h}).

\item A file called \texttt{strname.in} containing a white space\footnote{%
A \textquotedblleft white space\textquotedblright\ is a space, tab or newline.}%
-separated list of directories containing structures that will be used to
calculate force constants.\ Typically these would numbered directories
previously generated by \texttt{mmaps}, although user-specified directories
can be included. This set of structures must be diverse enough so that it
contains at least one chemical bond of each type that will be encountered
with the \texttt{svsl} code.\ For a binary A-B alloy this typically
involves\ the need for a \textquotedblleft pure A\textquotedblright\ and a
\textquotedblleft pure B\textquotedblright\ structure, in addition to one
with an intermediate composition.\ Each of the directories listed in \texttt{%
strname.in} must contain:

\begin{enumerate}
\item a \texttt{str.out} file containing an ideal unrelaxed structure that
will be used to automatically determine the nearest neighbor shell,

\item a \texttt{str\_relax.out} file containing the relaxed structure that
will be used to calculate bond lengths and that the code will perturb in
various ways to determine the force constants.
\end{enumerate}
\end{enumerate}

Like the \texttt{fitfc} code, the \texttt{fitsvsl} code can operate in two
modes: a structure generation mode and a fitting mode (indicated by the 
\texttt{-f} option). In structure generation mode, all the above input files
are needed and the option \texttt{-er} must be specified in order to
indicate the size of the supercells generated. The \texttt{-er} option
specifies the minimum required distance between a displaced atom's periodic
images.\ The code will automatically infer the smallest supercell satisfying
this constraint.\ Typically, \texttt{-er} should be 3 or 4 times the nearest
neighbor distance. All of these distances are measured using the ideal
structures (\texttt{str.out}).

Various optional parameters (which have reasonable default values) can be
specified: the \texttt{-dr}, \texttt{-ms} and \texttt{-ns} options have the
same meaning as for the \texttt{fitfc} code. Unlike the \texttt{fitfc} code,
the default number of different isotropic strain levels considered is 2
because this is the minimum needed to be able to determine the
length-dependence of bond stiffness. More than one level of strain is
necessary even if only the harmonic approximation is desired. Another
difference is that the \texttt{fitsvsl} code is typically run from a
\textquotedblleft top level\textquotedblright\ directory containing multiple
structure subdirectories (e.g. as generated by \texttt{mmaps}), because it
handles properties that are transferable across structures. The \texttt{fitfc%
} code is to be run within a structure directory because it applies to one
structure at the time.

After the structure generation step, each of the directory specified in 
\texttt{strname.in} will contain a hierarchy of subdirectories with various
perturbed structures with a layout\ identical to the one described in step %
\ref{fitfcstepfiles} of section \ref{secfitfc}\ above. The appropriate
first-principles calculations are then carried out as in step \ref%
{fitfcstepvasp} of that same section.

After completion of the first-principles calculations, \texttt{fitsvsl} must
be invoked in fitting mode (with the \texttt{-f} option). The lattice file
(e.g. \texttt{lat.in}) must be present and the code will look for data to
fit in all files of the form \texttt{*/force.out,\ */*/force.out} and the
corresponding files \texttt{*/str.out */str\_relax.out */*/str.out
*/*/str\_relax.out}.

The code will then use that information to create the length-dependent force
constants (this may take a few minutes) and ouputs them in \texttt{%
slspring.out}. Here is an example of the format of this file:

\begin{tabular}{ll}
\texttt{Al Al} & (gives the type of bond) \\ 
\texttt{2} & (2 parameters: linear fit is used) \\ 
& (polynomial coefficients of the stiffness vs length relationship for
stretching term:) \\ 
\texttt{50.29} & (constant, typically in eV/\AA $^{2}$) \\ 
\texttt{7.89} & (slope coefficient, typically in eV/\AA $^{3}$) \\ 
\texttt{2} & (idem for bending term, 2 parameters for a linear fit) \\ 
\texttt{6.13} & etc. \\ 
\texttt{-1.02} &  \\ 
\texttt{Al Ti} & (repeat for each type of chemical bond) \\ 
\ldots & 
\end{tabular}

Whenever there is not enough data to fit a given parameter, it will appear
as \texttt{0} in the above output file, so the user must carefully inspect
this file. This may happen if there are not enough structures in the \texttt{%
strname.in} file or if there are too few levels of strain applied (\texttt{%
-ns} option) or if some ab initio calculations aborted without producing a 
\texttt{force.out} file.

A few options are available to control the fitting process. The \texttt{-op}
option specifies the order of the polynomial used to fit the stiffness vs
length relationship (the default is \texttt{-op=1} for a linear
relationship). More sophisticated models\footnote{%
A more general format for the \texttt{slspring.out} file then used ---
contact the author for more information.} are also available, such as
direction-dependent force constants (\texttt{-dd} option) or
composition-dependent force constants \cite{liu:cdtfc} (the \texttt{-pc}
option specified the order of the polynomial representing the
composition-dependence).

\begin{figure}
\centerline{\epsfbox{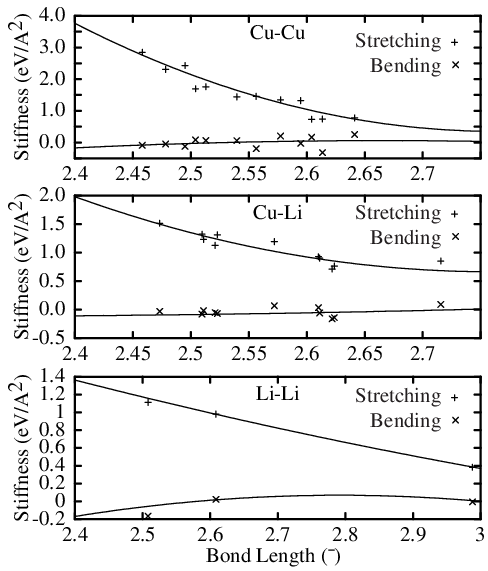}}
\caption{\label{figsvsl}
Bond stiffness versus bond length relationship in the fcc Cu-Li system.
The proximity of the crosses to the solid lines is a strong indication of the
model's predictive power in this system.
}
\end{figure}

Diagnostic files are also output: The \texttt{fitsvsl.log} file contains all
the matrices generated in the fitting process, including the actual forces,
compared with the ones predicted by the spring model. A plot of the
stiffness vs. length relationship is available in the files \texttt{%
fitsvsl.gnu} (a gnuplot script) and \texttt{f\_*.dat}\ the raw stiffness
data to be plotted. This is helpful for assessing the validity of the
transferability assumption, as illustrated in Figure \ref{figsvsl}.

Note that the \texttt{fitfc} and \texttt{fitsvsl} codes are mutually
compatible: The perturbations generated with one code can be used in a fit
carried out with the other. This is useful if one realizes midway through
the process that one approach is too slow or the other too inaccurate.

Once the \texttt{slspring.out} file has been generated with \texttt{fitsvsl}%
, the \texttt{svsl} code can exploit it to rapidly predict vibrational
properties of numerous similar structures. Like \texttt{fitfc}, \texttt{svsl}
is structure-specific and must be run within a structural directory
containing the following files (three of which should now be familiar):

\begin{enumerate}
\item a\ \textquotedblleft relaxed\textquotedblright\ structure file (%
\texttt{str\_relax.out}) and

\item an \textquotedblleft unrelaxed\textquotedblright\ structure file (%
\texttt{str.out}). The latter provides the ideal atomic positions that are
used to automatically determine which atoms lie in the nearest neighbor
shell. This file can be the same as the relaxed structure but then the
determination of the nearest neighbor shell may be less reliable.

\item In addition, \texttt{svsl} requires a transferable force constant file 
\texttt{slspring.out}. It looks for this file in the current directory and,
if not found, it also looks one directory up, where it would typically have
been placed by an earlier \texttt{fitsvsl} command.

\item An optional input file defining the atomic masses can be provided. By
default, the code looks in the \texttt{\symbol{126}/.atat.rc} file to
determine the directory where ATAT\ is installed and then loads the file 
\texttt{data/masses.in}. This behavior can be overridden with the \texttt{-m}
option.

\item By default, the bulk modulus is calculated from the force constants
(which is quick but potentially inaccurate) but it can also be read from a
file (whose name is specified with the \texttt{-bf} option) or specified on
the command line with the \texttt{-b} option.
\end{enumerate}

A few parameters govern the accuracy of the calculations.\ (The default
values are all reasonable starting points.)

\begin{enumerate}
\item By default the code uses the harmonic approximation, but thermal
expansion can be accounted for by specifying the \texttt{-ns} option with a
value $ns$ greater than 1. In that event, \texttt{svsl} calculates the
vibrational free for $ns$ different lattice parameters ranging from its
equilibrium 0K value to a value $(1+ms)$ times larger, where $ms$ is
specified via the \texttt{-ms} option (the default is 0.05). Typically 
\texttt{-ns=5} is sufficient, but the computational cost of using higher
values is minimal.\newline
The code calculates the vibrational free energy for temperatures ranging
from $T0$ to $T1$ in steps of $dT$, where these three values are determined
as follows.

\begin{enumerate}
\item The defaults are $T0=0$, $T1=2000$, $dT=100$.

\item If a \texttt{Trange.in} file exists in the next upper directory, it is
used to set $T0,T1,dT$: It must contain two numbers on one line separated by
a space: $T1$ $(T1/dT+1)$. Note that $T0=0$ always. For phase diagram
calculations, this method must be used to specify the temperature range
because it ensures that all calls to \texttt{svsl} (and to other codes
described subsequently) use the same temperature range.

\item The above values can be overridden by the \texttt{-T0=}$T0$, \texttt{%
-T1=}$T1$ and \texttt{-dT}=$dT$ options.
\end{enumerate}

\item The $k$-point sampling grid is specified with the \texttt{-kp}=$kp$
option. The actual number of $k$-points used is at least $kp$\ divided by
the number of atoms in the cell. An internal algorithm builds a $k$-point
grid that equalizes as much as possible the $k$-point density along all
directions in space
\end{enumerate}

The output files are as follows:

\begin{enumerate}
\item \texttt{svsl.log} : a log file giving some of the intermediate steps
of the calculations in particular, the predicted bulk modulus which should
be checked for accuracy.

\item \texttt{vdos.out}: the phonon density of states for each lattice
parameter considered (unstable modes appear as negative frequencies).

\item \texttt{svsl.out}: gives along each row, the temperature, the free
energy, and the linear thermal expansion (e.g. 0.01 means that the lattice
has expansion by 1\% at that temperature).

\item \texttt{fvib}: similar to \texttt{svsl.out} but gives only the free
energy.
\end{enumerate}

Examples of studies employing this code can be found in \cite%
{burton:AlGaInN,avdw:naclkcl,avdw:culi}

\subsection{Electronic excitations}

The \texttt{felec} code calculates the electronic free energy within the
one-electron and temperature-independent bands approximations. It expects a 
\texttt{dos.out} input file containing an electronic density of states and
the number of electrons to populate it. This file is typically generated by
the ab initio call up utilities \texttt{runstuct\_xxxx} (see Section \ref%
{secabi}). It is rarely needed to use anything but the default option:%
\newline
\hspace*{0.5in}%
\texttt{felec -d}

The output files are:

\begin{enumerate}
\item \texttt{felec.log}, containing the temperature and corresponding free
energy (per unit cell) on each line.

\item \texttt{felec} is similar to \texttt{felec.log} but only contains the
free energies.

\item \texttt{plotdos.out} contains the dos in a format suitable to be
plotted, with the energy normalized so that the Fermi level is at 0.
\end{enumerate}

It should be noted that the above approximation are not always satisfactory,
notably in magnetic systems. There are two (currently experimental) magnetic
free energy calculators included in ATAT:

\begin{enumerate}
\item \texttt{fmag}, which relies on a simple quantum mechanical mean field
model and

\item \texttt{fempmag}, which implements the widely used semi-empirical
method introduced in \cite{ohnuma:empmag}.
\end{enumerate}

\subsection{Cluster expansions of nonconfigurational free energies}

\label{seccenc}The \texttt{fitfc}, \texttt{svsl} and \texttt{felec} codes
enables the calculation of free energies of individual configuration on a
lattice. However, to properly calculate the free energy of system exhibiting
both configurational and nonconfigurational disorder, a cluster expansion
needs to be constructed to be able to predict the nonconfigurational free
energy of \emph{any} given configuration. This cluster expansion can then be
provided as an input to the Monte Carlo code \texttt{memc2} which will now
be able to properly account for both configurational and nonconfigurational
contributions to the free energy. Here is how to proceed.

\begin{enumerate}
\item First create a \texttt{Trange.in} file indicating the temperature
range to be sampled. For instance, if one wishes to sample from 0K to 2000K
in intervals of 100K, type\newline
\hspace*{0.5in}%
\texttt{echo 2000 21 {>} Trange.in}\newline
(The lowest temperature is always 0K and note that $2000/(21-1)=100$.)%
\newline
This ensures that all calls to \texttt{fitfc}, \texttt{svsl} and \texttt{%
felec} with use the same temperature range.

\item ATAT provides the \texttt{foreachfile} utility to repeat the same
command in all subdirectories containing a specified file. For instance, to
calculate the vibrational free energy using \texttt{svsl} in all structure
subdirectories, one could type\newline
\hspace*{0.5in}%
\texttt{foreachfile -e str\_relax.out svsl -d}\newline
This executes the \texttt{svsl} code with the default options (\texttt{-d})
in each subdirectory containing a \texttt{str\_relax.out} file while
skipping directories with an\ \texttt{error} file (\texttt{-e}). If desired,
a similar process can be used for the electronic free energies:\newline
\texttt{foreachfile -e dos.out felec -d}

\item Next, construct a cluster expansion of the vibrational free energy
data contained in the \texttt{*/fvib} files:\newline
\hspace*{0.5in}%
\texttt{clusterexpand fvib}\newline
Note that the files \texttt{*/fvib} actually contain a column of numbers
(one for each temperature) --- \texttt{clusterexpand} simply produce a
separate cluster expansion for each scalar element in \texttt{*/fvib}.\ By
default, this commands reads in the clusters from the \texttt{clusters.out}
file, as they were generated earlier by \texttt{mmaps}.\footnote{%
If the user wishes to use \texttt{clusterexpand} without first running 
\texttt{mmaps}, a suitable \texttt{clusters.out} file can be generated with
the a command of the form \texttt{corrdump -clus -2=[max pair radius]
-3=[max triplet radius]} etc. A \texttt{lat.in} file still needs to be
supplied.} The user has some control over which clusters to include\ (1) or
exclude (0) via the \texttt{-s=[comma-separated string of 0s and 1s]}
option. The user can also check if the crossvalidation score, reported with 
\texttt{-cv} option, can be improved by changing the cluster selection. A
similar expansion can be done for electronic contributions:\newline
\hspace*{0.5in}%
\texttt{clusterexpand felec}\newline

\item Finally, combine the energy cluster expansion from \texttt{eci.out}
(generated by \texttt{mmaps}) with the vibrational (\texttt{fvib.eci}) and
electronic (\texttt{felec.eci}) cluster expansions created in the previous
step into a single \textquotedblleft master\textquotedblright\ cluster
expansion file (\texttt{teci.out}) that will be read by the Monte Carlo code:%
\newline
\hspace*{0.5in}%
\texttt{mkteci fvib.eci felec.eci}\newline
For this step to succeed it is essential that all \texttt{clusterexpand}
commands were run with the same \texttt{mmaps}-generated \texttt{clusters.out%
} file (although different clusters can be included or excluded with the 
\texttt{-s} option of \texttt{clusterexpand} without any problem).

\item Run \texttt{memc2}. It is often instructive to try and compare
simulations that include or exclude some components of the
nonconfigurational free energy. To this effect, one can simply rerun \texttt{%
mkteci} with the list of contributions to include before rerunning \texttt{%
memc2}.
\end{enumerate}

Example of the use of this methodology can be found in \cite%
{burton:AlGaInN,avdw:naclkcl,avdw:culi}. For easy reference, Figure \ref%
{figflow} shows a flowchart of the entire procedure while Figure \ref%
{figculi} showcases a practical example of this methodology.

\begin{figure}
\centerline{\epsfbox{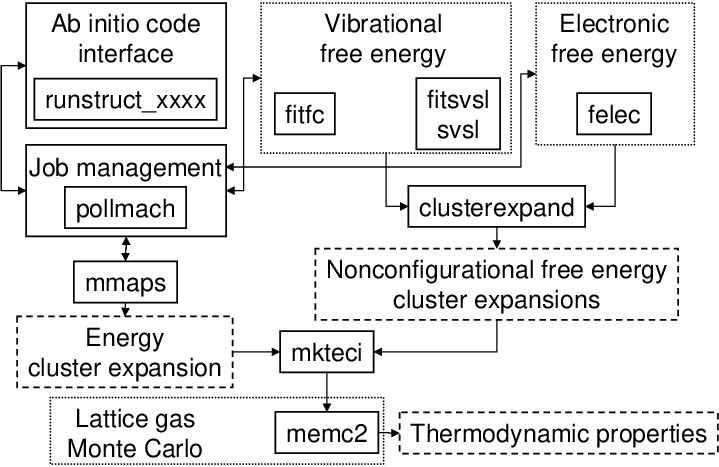}}
\caption{\label{figflow}Flowchart of ATAT's approach to the integration of
configurational and nonconfigurational contributions to the thermodynamics of alloys.}
\end{figure}%

\begin{figure}
\centerline{\epsfbox{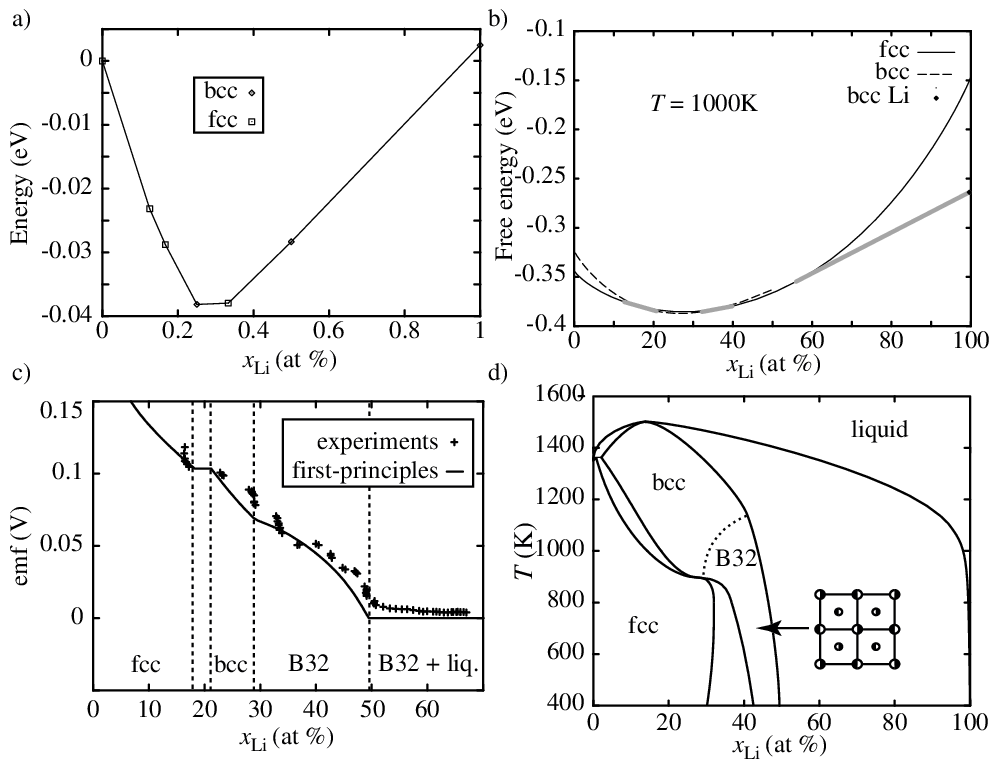}}
\caption{\label{figculi}
Ab initio thermodynamics of the Cu-Li system \cite{avdw:culi}.
a) Ground state convex hull (merging bcc and fcc superstructures).
b) Calculated free energy as a function of composition including both configurational,
vibrational and electronic contributions. Nonconfigurational contributions are essential
to accurately predict the numerous phase transitions between bcc- and fcc-based phases.
Two-phase equilibria are marked by thick gray tie-lines.
c) Calculated electromotive force (emf) inferred from the calculated free energy in b)
and compared with experimental data. (The solid-liquid Li free energy difference was
left as a parameter adjusted to provide the best agreement.)
d) Calculated phase diagram obtained via tangent constuctions as in b) over a
range of temperatures. The liquid phase excess free enegy is taken from the
COST database \cite{saunders:costdb}. Note that most of the ground states from a)
disorder at low temperature and are not visible in the phase diagram. Interestingly,
neither the magnitude of each ground state's formation energy nor the sharpness of the
vertex where a ground state touches the hull bear any correlation with a ground state's
disordering temperature. The B32 ground state at 50\% (shown in inset) does not have the
most negative formation energy and barely breaks the hull. Yet it exhibits
tremendous stability due to entropy-driven stabilization. The dashed line marks the
location of a second-order transition, determined by locating a peak in the heat capacity.
}
\end{figure}%

\section{Special Quasirandom Structure generation}

The use of the cluster expansion is essential to properly model nondilute
alloys exhibiting configurational disorder with potential short-range order (%
\cite{ghosh:sqsce}). However, at sufficiently high temperatures, the
limiting case of a fully random alloy provides a useful approximation.
Although this limit can be calculated with a cluster expansion, it is
sometimes convenient to directly determine high-temperature limiting
quantities with a small number of supercell ab initio calculations without
first constructing a cluster expansion. The concept of Special Quasirandom
Structures (SQS) \cite{zunger:sqs}\ provides an efficient way to take this
route. An SQS represents the best periodic supercell approximation to the
true disordered state for a given number of atoms per supercell. The quality
of an SQS is measured in terms of the number of correlations of the fully
disordered state it is able to match exactly. Typically, one attempts to
preferably match shorter-range correlations while gradually enlarging the
supercell to extend the range of matching correlations until convergence of
the properties of interest. Traditionally, SQS are built based on matching
the pair correlations, although for best performances, multibody
correlations should be considered as well. The SQS generation utility
included in ATAT (\texttt{gensqs}) can account for any user-specified set of
correlations.

The SQS generation process in ATAT (as used, for instance, in \cite%
{ghosh:sqsce,shin:hcpsqs,shin:ternsqs}) is as follows:

\begin{enumerate}
\item Prepare a \texttt{lat.in} file (just like the one used for the \texttt{%
mmaps} and \texttt{memc2} codes).

\item Prepare a file (say, \texttt{conc.in}) specifying\ a structure having
the same overall composition as\ the disordered state of interest. The exact
geometry of this structure is irrelevant --- it only needs to have the
correct overall composition and be a proper superstructure of \texttt{lat.in}%
. Each site must be occupied by one and only one atom. For instance, given
the lattice of Table \ref{tablat}, a valid \texttt{conc.in} file could be:%
\newline
\hspace*{0.5in}%
\begin{tabular}{l}
\texttt{1 1 1.05 90 90 90} \\ 
\texttt{1.0 \ 0.0 \ 0.0} \\ 
\texttt{0.0 \ 1.0 \ 0.0} \\ 
\texttt{0.0 \ 0.0 1.0} \\ 
\texttt{0 \ \ 0 \ \ 0 \ \ Fe} \\ 
\texttt{0 \ \ 0.5 0.5 C} \\ 
\texttt{0.5 0 \ \ 0.5 C} \\ 
\texttt{0.5\ 0.5 0 \ \ C,Vac} \\ 
\texttt{0.5\ 0.5\ 0.5\ Ni} \\ 
\texttt{0.5\ 1.0 1.0 C,Vac} \\ 
\texttt{1.0 0.5 1.0 C,Vac} \\ 
\texttt{1.0\ 1.0 0.5\ C,Vac}%
\end{tabular}%
\newline
This way of specifying composition was selected because it allows the user
to individually control the composition on each sublattice.

\item Select a small number of correlations to match exactly and compute
their value in the disordered state:\newline
\hspace*{0.5in}%
\texttt{corrdump -noe -rnd -s=conc.in -2=[pair max radius] --3=[triplet max
radius] {>} tcorr.out}\newline
This command writes the nonempty (\texttt{-noe}) correlations of the
disordered state (\texttt{-rnd}) having the same composition as the
structure \texttt{conc.in} into the file \texttt{tcorr.out}. Pair
correlations up to distance \texttt{[pair max radius]} and triplet
correlations up to \texttt{[pair triplet radius]} are considered (larger
clusters can also be specified with \texttt{-4=\ldots }, etc.). This command
also writes a cluster file \texttt{clusters.out} that will be read in the
step.

\item The SQS generation step is initiated with, for instance,\newline
\hspace*{0.5in}%
\texttt{gensqs -n=16 {>} sqs16.out}\newline
which generates 16-atom SQS and puts them into the \texttt{sqs16.out} file.
If this file is empty, the range of correlations to be matched exactly needs
to be reduced in step 3. A word of caution: \texttt{gensqs} only generates
structures containing exactly the number of atoms per unit cell specified by
the \texttt{-n} option (i.e., if an SQS with a smaller unit cell exists, it
will not be listed).

\item The output file \texttt{sqs16.out} may contain a large number of
candidate SQS. While these a match the specified criteria, they may differ
in quality in terms of the correlations not included in the screening
process. Further screening can be performed with the command\newline
\hspace*{0.5in}%
\texttt{corrdump -2=[another radius] -3=[another radius] -4=[another radius]
-noe -s=sqs16.out}\newline
which will output the correlations of all SQS in \texttt{sqs16.out}.
Clearly, the radii specified should be larger than in step 3 in order to
distinguish and rank these SQS.
\end{enumerate}

While there are no formal rules to decide whether it is preferable to match
more pairs or more multi-body correlations, general guidelines can be given.
In general, it would be unusual to match longer range $n$-body correlations
than $m$-body interactions if $n>m$. In systems exhibiting large relaxations
(due to large atomic size mismatches), multi-body correlations are expected
be important, while in ionic systems, pair interactions are expected to
dominate.

SQS generation via exhaustive enumeration is time-consuming. This can be
alleviated by exploiting parallelism. Multiple independent copies of \texttt{%
gensqs} can be run on separate processors if one specifies the \texttt{%
-ip=[index]} and \texttt{-tp=[number of processes]} options. The \texttt{%
[index]} runs from 0 to \texttt{[number of processes-1]} and is used to tell
each copy of \texttt{gensqs} which portion of the space to scan. For maximum
efficiency, \texttt{[number of processes]} should be a factor of the total
number of possible supercells, which can be determined, e.g., with the 
\texttt{gensqs -n=16 -pc} command.

Another way to improve SQS generation speed is via stochastic sampling \cite%
{abrikosov:sqsmc}. Such an approach seeks to quickly find close-to-optimal
solutions rather than slowly finding an exact solution. Another approach is
to scan through configurations in such as way that only the desired
composition is sampled, using atom permutations rather than atom
\textquotedblleft transmutations\textquotedblright . This scheme becomes
increasingly advantageous as one moves away from equiatomic compositions. Yi
Wang contributed a useful code implementing these two algorithms (that were
used in \cite{shin:ternsqs}), which will be included in the ATAT distibution
shortly.

Finally, we note that SQS prove useful in the construction of a cluster
expansion as well. Any generated SQS can be included into the set structures
used to fit a cluster expansion as a way to ensure that the properties of
the disordered state are properly reproduced. In practice, one can merely
place the desired SQS into a \texttt{str.out} file within a subdirectory and
run an ab initio code on it with \texttt{runstruct\_xxxx}. The \texttt{mmaps}
code will readily read in any user-specified structure placed in its
directory hierarchy (provided the \texttt{lat.in} files used in \texttt{mmaps%
} and \texttt{gensqs} are identical).

\section{The tensorial cluster expansion}

While the formalism enabling the cluster expansion of tensors has been
derived in \cite{avdw:gce}, we provide here a brief explanation of how to
use this feature. In Section \ref{seccenc}, it was shown how to cluster
expand a scalar --- the process is similar for tensors.

\begin{enumerate}
\item Create a \texttt{lat.in} providing the lattice and a \texttt{%
gcetensor.in} file describing the type of tensor in the form:\newline
\hspace*{0.5in}%
\texttt{[rank]}\newline
\hspace*{0.5in}%
\texttt{[list of pairs of indices indicating which simultaneous index
permutations leave the tensor invariant]}\newline
\hspace*{0.5in}%
\texttt{[next list, etc...]}\newline
To fix the ideas, here are a few examples. For strain or stress the \texttt{%
gcetensor.in} should be:\newline
\hspace*{0.5in}%
\texttt{2}\newline
\hspace*{0.5in}%
\texttt{0 1} (indicating that such tensor is symmetric, i.e.,\ $\varepsilon
_{ij}=\varepsilon _{ji}$)\newline
while for elastic constants, it should be:\newline
\hspace*{0.5in}
\texttt{4}\newline
\hspace*{0.5in}
\texttt{0 1} (indicating $c_{ijkl}=c_{jikl}$)\newline
\hspace*{0.5in}
\texttt{2 3} (indicating $c_{ijkl}=c_{ijlk}$)\newline
\hspace*{0.5in}
\texttt{0 2 1 3} (indicating $c_{ijkl}=c_{klij}$).

\item Generate a list of clusters with the command\newline
\hspace*{0.5in}%
\texttt{gce -clus -2=[max pair radius] -3=[max triplet radius]} etc.\newline
Note that \texttt{gce} (which stands for Generalized Cluster Expansion)
admits essentially the same syntax as \texttt{corrdump} in the scalar case.

\item Calculate the property tensor associated with each structure.\footnote{%
This list of structures could come from a previous \texttt{mmaps} run, for
instance, or it could be generated manually using the \texttt{genstr}
command.} This is highly application-dependent. For instance, to calculate
the static lattice strain, one could use (assuming ab initio calculations
have been run in each subdirectory):\newline
\hspace*{0.5in}%
\texttt{foreachfile -e str\_relax.out analrelax -d "{>}" strain}%
\newline
To calculate the elastic constants one could use:\newline
\hspace*{0.5in}%
\texttt{foreachfile -e str\_relax.out calcelas -d} (this generates
perturbations)\newline
\hspace*{0.5in}%
\texttt{pollmach -e runstruct\_xxxx -w force.wrap} (this calculates the
induced reaction stresses)\newline
\hspace*{0.5in}%
\texttt{foreachfile -e str\_relax.out calcelas -f "{>}" elas}
(this fits the elastic tensor and writes it in the file \texttt{elas}.)

\item Finally, do the cluster expansion itself with:\newline
\hspace*{0.5in}%
\texttt{clusterexpand -pa -g strain}\newline
\hspace*{0.5in}%
\texttt{clusterexpand -pa -g elas}\newline
the \texttt{-g} invokes the generalized cluster expansion option while the 
\texttt{-pa} indicates that the quantities to be expanded are intensive
(i.e. per atom).
\end{enumerate}

\section{Interfaces with ab initio codes}

\label{secabi}All interfaces have the general name \texttt{runstruct\_xxxx}.
Currently, the following are supported: \texttt{runstruct\_vasp} (for the
VASP code \cite{kresse:vasp1,kresse:vasp2}) \texttt{runstruct\_abinit} (for
the abinit \cite{gonze:abinit1,gonze:abinit2}\ code), \texttt{runstruct\_gulp%
} (for the GULP code \cite{gale:gulp,gale:gulpsym}). In addition, Monodeep
Chakraborty, J\"{u}rgen Spitaler, Peter Puschnig and Claudia Ambrosch-Draxl
have written an interface with \texttt{WIEN2k} \cite{blaha:wien2k}, which
will be publicly available shortly. Finally, \texttt{runstruct\_pwscf} (for
the PWscf code \cite{giannozzi:pwscf}) is under development.

Each interface\ has the following characteristics in common:

\begin{enumerate}
\item It reads the geometry of a structure from the \texttt{str.out} file in
current directory, written in the ATAT format. If a \texttt{str\_hint.out}
file exists, it takes precedence over the \texttt{str.out} file. This is
useful to provide educated\ guesses of the relaxed geometry via external
user-supplied codes.

\item It reads some code-specific parameters from a file called \texttt{%
xxxx.wrap} (or specified with \texttt{-w [file]}) located in the current
directory or up to two levels above. This separation between the geometry
and calculation parameter input files is essential to ensure that all the
pieces of ATAT are fully interoperable.

\item It runs the appropriate ab initio or atomistic code. If an argument is
given to the \texttt{runstruct\_xxxx} this string is used as prefix to the
ab initio command. This feature is used to run the appropriate code remotely
without having to install ATAT on the compute nodes.

\item It writes the structure's energy in the file \texttt{energy} and the
structure's relaxed geometry in \texttt{str\_relax.out}. (If no relaxations
are allowed, then the unrelaxed geometry is written in \texttt{str\_relax.out%
}.)

\item It writes the forces acting on all atoms in \texttt{forces.out}, the
stress acting on the cell in \texttt{stress.out} and the electronic density
of states in \texttt{dos.out} (although these are not yet implemented in all 
\texttt{runstruct\_xxxx} commands).

\item If anything goes wrong with the calculations, a file called \texttt{%
error} is written.
\end{enumerate}

Typically the ab initio codes are not called one run at the time but rather
as a batch of many jobs. The \texttt{pollmach} command manages such pools of
jobs. The basic principle is simple: If a file called \texttt{wait} exists
in a subdirectory, \texttt{pollmach} will find it and run the command
specified on the command line within that directory. For instance,\newline
\hspace*{0.5in}%
\texttt{pollmach runstruct\_xxxx}\newline
will repeatedly wait for a \texttt{wait} file to be found, delete it, and
then run \texttt{runstruct\_xxxx} with the corresponding directory. In a
multiprocessor environment, \texttt{pollmach} can simultaneously dispatch
different jobs to different processors. This simple mechanism allows most of
the ATAT codes to be completely platform-independent, with \texttt{pollmach}
being the only platform-dependent piece.

The first time \texttt{pollmach} is run, default configurational files are
set up and the user will be asked to tailor them to the local computing
environment. The ATAT distribution includes many examples of setups,
including the increasingly common case of a large pool of processors to be
divided up into subgroup that each run a separate parallel version of an ab
initio code.

\section{Utilities}

We now briefly mention some utilities to give users direct access to some of
the inner algorithms of ATAT.

\begin{enumerate}
\item \texttt{corrdump} finds symmetry operations, enumerates clusters,
calculates correlations (including of the disordered state), etc.

\item \texttt{genstr} enumerates superstructures of a given lattice.

\item \texttt{pdef} generates substitutional point defect supercells.

\item \texttt{cellcvrt} manipulates ATAT-formatted structure files, changing
coordinate systems, converting fractional to cartesian coordinates, finding
supercells and subcells, etc.

\item \texttt{lsfit} implements least-squares fitting.

\item \texttt{nnshell} finds nearest-neighbor shells.

\item A number of text parsing utilities: \texttt{getvalue}, \texttt{getlines%
}, \texttt{sspp}.
\end{enumerate}

More information regarding these utilities can be found in their respective
help files (accessed by specifying the \texttt{-h} option).

\section{Conclusion}

This completes this overview of the various new features added to ATAT in
recent years. What's next? Probably:

\begin{enumerate}
\item A tighter integration between the output of the Monte Carlo code and
thermodynamic databases for use with softwares such as Thermocalc and Pandat;

\item Material propery optimization modules exploiting the tensorial cluster
expansion;

\item More automated ways to include nonconfigurational free energy;

\item Better electronic free energy calculators.
\end{enumerate}

However, in large part, what will be next will depend on what users express
interest in and what the author can get funding for\ldots

\section*{Acknowledgements}

This research was supported by the US National Science Foundation through
TeraGrid resources provided by NCSA and SDSC under grant TG-DMR050013N. The
NSF Center for the Science and Engineering of Materials at Caltech supported
the preparation of this manuscript.

The author would like to thank

\begin{itemize}
\item Mark Asta for supporting this effort while the author was at
Northwestern.

\item Gerd Ceder for supporting this effort while the author was at MIT.

\item Yi Wang, who has contributed an efficient SQS generator.

\item Dongwong Shin, who has contributed a list of useful lattices.

\item Volker Blum, who provided a nice perl remake of the load checking
utility \texttt{chl}.

\item Mayeul d'Avezac, who fixed a number of tricky compilation problems and
identified a few bugs.

\item Gautam Ghosh, who provided a utility to convert \texttt{emc2} output
files into a format suitable for the Thermocalc's Parrot module. He also
suggested the inclusion of magnetic free energies based on \cite%
{ohnuma:empmag}.

\item Zhe Liu for his contributions to generalize the Stiffness vs. Length
approach to include composition-dependence.

\item Monodeep Chakraborty, J\"{u}rgen Spitaler, Peter Puschnig and Claudia
Ambrosch-Draxl, who are developing an interface with \texttt{WIEN2k}.

\item Ben Burton and Raymundo Arroyave, who helped debug the code and the
documentation.

\item Pratyush Tiwary, who proofread this manual.
\end{itemize}

These faithful and patient users (and many others) have been crucial to help
maintain and develop this toolkit.

\appendix

\section{Correlation to concentration conversions}

\label{appconv}

ATAT calculates the matrices $C$\ and $X$ and vectors $c_{0}$ and $x_{0}$ in
Equations (\ref{eqgivec}) and (\ref{eqgivex}) as follows.

\begin{enumerate}
\item Start enumerating all structures (in order of increasing unit cell
size) while computing the point correlation vector $\rho $ for each of them.
Let $\bar{\rho}$ denote the corresponding point correlation vector with a
constant element appended to it.\ Skip any structures whose $\bar{\rho}$ is
colinear with the $\bar{\rho}$ of earlier structures. Terminate this step
when the number of structures kept is equal to the dimension of $\bar{\rho}$.

\item For each of the kept structures, calculate its full concentration
vector $x$. Create a matrix $A$ by pasting the column vectors $x$ just
obtained side by side. Similar, create a matrix $B$ by pasting the column
vectors $\bar{\rho}$ previously obtained. Note that $B$ is square and
invertible by construction and calculate $\bar{X}=AB^{-1}$.

\item All columns of $\bar{X}$ but the last provide $X$ while the last
column of $\bar{X}$ gives $x_{0}$.

\item Eliminating all colinear rows from $X$ gives $C$ while eliminating the
corresponding elements of $x_{0}$ gives $c_{0}$. (Note: in ATAT, $c_{0}$ is
just set to zero since this makes no difference for convex hull
construction. But $x_{0}$ is fully calculated.)
\end{enumerate}

\end{document}